\documentclass[aps,pra,twocolumn,superscriptaddress,showpacs]{revtex4}
\usepackage{amsfonts}
\usepackage{amsmath}
\usepackage{amssymb}
\usepackage{graphicx}
\usepackage[english]{babel}
\usepackage{hyperref}
\usepackage[usenames,dvipsnames]{color}
\usepackage{babel}

\makeatletter
\begin{document}

\title{Speeding up antidynamical Casimir effect with nonstationary qutrits}
\author{A. V. Dodonov }
\email{adodonov@fis.unb.br}
\affiliation{
Institute of Physics and International Centre for Condensed Matter Physics, University of Brasilia, 70910-900, Brasilia, Federal District, Brazil}
\author{J. J. D\'iaz-Guevara}
\affiliation{Departamento de F\'isica, Universidad de Guadalajara, Revoluci\'on 1500,
Guadalajara, Jalisco 44420, Mexico}
\author{A. Napoli}
\affiliation{Dipartimento di Fisica e Chimica, Universit\`a degli Studi di Palermo, Via Archirafi 36, I-90123 Palermo, Italy}
\affiliation{I.N.F.N. Sezione di Catania}
\author{B. Militello}
\affiliation{Dipartimento di Fisica e Chimica, Universit\`a degli Studi di Palermo, Via Archirafi 36, I-90123 Palermo, Italy}
\affiliation{I.N.F.N. Sezione di Catania}

\begin{abstract}
The antidynamical Casimir effect (ADCE) is a term coined to designate the
coherent annihilation of excitations due to resonant external perturbation
of system parameters, allowing for extraction of quantum work from
nonvacuum states of some field. Originally proposed for a two-level atom (qubit) coupled to
a single cavity mode in the context of nonstationary quantum Rabi model, it suffered from
very low transition rate and correspondingly narrow resonance linewidth. In
this paper we show analytically and numerically that the ADCE rate can be
increased by at least one order of magnitude by replacing the qubit by an
artificial three-level atom (qutrit) in a properly chosen configuration. For the cavity thermal state we demonstrate that the dynamics of the average photon number and atomic excitation is completely different from the qubit's case, while the behavior of the total number of excitations is qualitatively similar yet significantly faster.
\end{abstract}


\maketitle

\newcommand{\remove}[1]{{\it\color{red} #1}}

\section{Introduction}

The broad term \emph{dynamical Casimir effect} (DCE) refers to the
generation of excitations of some field (Electromagnetic, in the majority of
cases) due to time-dependent boundary conditions, such as changes in the
geometry or material properties of the system \cite{moore,fulling,dce1,dce2} (see \cite{dce-rev1,dce-rev2} for reviews; see also \cite{movbnd1,movbnd2,movbnd3} for the related problem of a particle in a wall with moving boundaries). In the so called \emph{cavity DCE} one
considers nonadiabatic (periodic or not) modulation of the cavity natural
frequency by an external agent, investigating the accumulation of intracavity
photons or the photon emission outside the cavity \cite{moore,law,lambrecht,zeilinger}. The additional interaction of the cavity
field with a stationary `detector' during the modulation (harmonic oscillator, few-level
atom or a set of two-level atoms in the simplest examples) may dramatically alter
the photon generation dynamics, for instance, altering the
field statistics, shifting the resonance frequency and inhibiting the photon
growth \cite{pla,naroz,roberto,dodo11-pla,2level,3level,cache} (see \cite{camop} for a
short review). Moreover, the degree of excitation of the detector varies
according to the regime of parameters, and entanglement can be created
between the cavity field and the detector, or between the set of atoms
coupled to the field \cite{2atom,Nlevel,solano1,stassi,rosatto}.

Over the past ten years a new path has attracted attention of the community
working on nonstationary phenomena in cavity Quantum Electrodynamics (QED).
Instead of changing the cavity frequency, different studies suggested the
parametric modulation of the `detector' instead, promoting it from a passive
to an active agent responsible for both the generation and detection of
photons \cite{jpcs,liberato,red1,red2,vedral,jpa,blais-exp,benenti,igor,palermo}. Beside eliminating the inconvenience of time-dependent Fock states
of the field associated to time-varying cavity frequency \cite{law}, this scheme makes
full use of the counter-rotating terms in the light--matter interaction
Hamiltonian and does not require the inclusion of additional parametric
down-conversion terms in the formalism \cite{jpcs,jpa,igor,palermo}.
Moreover, it benefits from recent advances in the coherent control and
readout of microscopic few-level quantum devices developed in the realm of
the circuit QED for applications in Quantum Information Processing (see \cite{nori2017} for a recent review).

The area of circuit QED investigates the interaction of artificial
superconducting atoms, formed by a sophisticated array of Josephson
Junctions, and the Electromagnetic field confined in increasingly complex
microwave resonators, ranging from waveguide resonators or 3D cavities \cite{pra-blais,c1,transmon,c2,apl}. The advances in engineering allowed for
implementation of multi-level atoms, with controllable transition
frequencies and coupling strengths, that can interact with multiple cavities
and other atoms controlled independently \cite{c1,blais-exp,m4,ge,ger,m1,m2,m3,nori2017,paik}. Moreover, circuit QED allows for
unprecedented atom--field coupling strength, in what became known as
ultrastrong and deep strong coupling regimes \cite{u1,u4,u2,u3}. In the context of DCE, the
exquisite control over the parameters of the Hamiltonian allows for
multi-tone multi-parameter modulations \cite{jpcs,porras,diego,tiago}, while quantum optimal control
strategies can be used to enhance the desired effects \cite{control}.

Photon generation is not the only phenomenon induced by parametric
modulations in circuit QED. It was shown recently that the counter-rotating
terms can also be employed to annihilate excitations of the Electromagnetic field from nonvacuum
initial states, in what became known as \emph{antidynamical Casimir effect}
(ADCE) \cite{igor}. This effect was predicted in the context of the quantum Rabi model, which describes the interaction of the
cavity field with a two-level atom \cite{rabi1,rabi2,rabi3}, and consists in the coherent
annihilation of three
photons accompanied by the excitation of the far-detuned atom \cite{lucas,tom} (four photons could be annihilated by employing a two-tone modulation \cite{diego}) .
Thus an amount of energy $\lesssim 2\hbar \omega _{0}$ could be extracted from the
system due to resonant perturbation of some parameter, where $\omega _{0}$
is the cavity frequency \cite{tiago}. However, in the more accessible regime of
weak atom--field interaction (beneath the ultrastrong coupling regime)
the associated transition rate is quite small, so the modulation frequency
must be finely tuned and the dissipation strongly affects the behavior \cite{diego,tiago}.

In this paper we uncover that the ADCE rate can be enhanced by almost two
orders of magnitude by employing artificial three-level atoms (qutrits) in
the standard ladder configuration and weak coupling regime \cite{transmon}. We
obtain closed approximate description of the unitary dynamics when one or more atomic parameters undergo a low-amplitude multi-tone external perturbation, and assess the
advantages and disadvantages of different regimes of parameters for the
initial thermal state of the cavity field. We also discuss eventual
complications that qutrits bring into the problem, such as adjustment of atomic energy levels with respect to the cavity frequency and two-tone driving with management of the modulation phases. Nevertheless it is argued that the
substantial gain in the ADCE rate compensates for the additional technical issues.

This paper is organized as follows. In Sec. \ref{sec1} we define our
problem and derive the general mathematical formalism to obtain approximate
expressions for the system dynamics in the dressed-states basis. In Sec. \ref{sec2} we discuss three
specific configurations of the qutrit for which the overall behavior is most easily inferred: the double-resonant, dispersive and mixed regimes. In
Sec. \ref{sec3} we identify the regimes of parameters and the transitions
for which excitations can be annihilated from the cavity thermal state,
assuming that the atom was initially in the lowest energy state. In
Sec. \ref{numerical} we evaluate
analytically the transition rates associated to ADCE between different
dressed states and compare our predictions to the exact numerical solution of the Schr\"{o}dinger equation, demonstrating that the ADCE rate can undergo almost
50-fold increase compared to the qubit's case while the amount of annihilated
excitations is roughly the same. Our conclusions are summarized in Sec. \ref{conclusions}.

\section{Mathematical formalism\label{sec1}}

We consider a three-level artificial atom (qutrit) interacting with a single
cavity mode of constant frequency $\omega _{0}$, as described by the
Hamiltonian (we set $\hbar =1$)%
\begin{equation}
\hat{H}=\omega _{0}\hat{n}+\sum_{k=1}^{2}E_{k}\hat{\sigma}%
_{k,k}+\sum_{k=0}^{1}G_{k}(\hat{a}+\hat{a}^{\dagger })(\hat{\sigma}_{k+1,k}+%
\hat{\sigma}_{k,k+1}).  \label{H2}
\end{equation}%
$\hat{a}$ ($\hat{a}^{\dagger }$) is the cavity annihilation (creation)
operator and $\hat{n}=\hat{a}^{\dagger }\hat{a}$ is the photon number
operator. The atomic eigenenergies are $E_{0}\equiv 0,E_{1}$ and $E_{2}$,
with the corresponding states denoted as $|\mathbf{0}\rangle ,|\mathbf{1}%
\rangle ,|\mathbf{2}\rangle $; the atomic operators read $\hat{\sigma}%
_{k,j}\equiv |\mathbf{k}\rangle \langle \mathbf{j}|$. The parameters $G_{k}$
($k=0,1$) stand for the coupling strengths between the atomic states $\{|%
\mathbf{k}\rangle ,|\mathbf{k+1}\rangle \}$ mediated by the cavity field.

We assume that all the atomic parameters can be modulated externally as 
\begin{equation*}
E_{k}(t)\equiv E_{0,k} + \varepsilon _{E,k}f_{E,k}(t),\quad G_{k}(t)\equiv
G_{0,k}+\varepsilon _{G,k}f_{G,k}(t),
\end{equation*} 
where $\{\varepsilon
_{E,k},\varepsilon _{G,k}\}$ are the modulation depths and $%
\{E_{0,k},G_{0,k}\}$ are the corresponding bare values. The dimensionless
functions%
\begin{equation}
f_{l}(t)=\sum_{j}w_{l}^{(j)}\sin \left( \eta ^{(j)}t+\phi _{l}^{(j)}\right)
\end{equation}%
represent the externally prescribed modulation, where the collective index $%
l $ denotes $\{E;k=1,2\}$ or $\{G;k=0,1\}$. Constants $0\le w_{l}^{(j)}\leq 1$
and $\phi _{l}^{(j)}$ are the weight and the phase corresponding to the
harmonic modulation of $l$ with frequency $\eta ^{(j)}$, and the index $j$ runs
over all the imposed frequencies (in this paper at most $2$-tone modulations will be examined). We normalize the weights so that $%
\sum_{j}w_{l}^{(j)}=1$ for any set $l$, so that $\varepsilon _{l}$
characterizes completely the modulation strength (in our examples we shall set $w_{l}^{(j)}=1$ and $\phi _{l}^{(j)}=0$ unless stated otherwise).

To obtain a closed analytical description we first rewrite the Hamiltonian
as $\hat{H}=\hat{H}_{0}+\hat{H}_{c}$, where%
\begin{equation}
\hat{H}_{0}=\omega _{0}\hat{n}+\sum_{k=0}^{2}\left[ E_{0,k}\hat{\sigma}%
_{k,k}+G_{0,k}(\hat{a}\hat{\sigma}_{k+1,k}+\hat{a}^{\dagger }\hat{\sigma}%
_{k,k+1})\right]
\end{equation}%
is the bare Hamiltonian in the absence of modulation and counter-rotating
terms (to shorten the formulas we defined formally $G_{0,2}=\varepsilon
_{G,2}=0$). For the realistic \emph{weak coupling} regime ($G_{0,0},G_{0,1}\ll \omega _{0}$) we
expand the wavefunction corresponding to the total Hamiltonian $\hat{H}$ as%
\begin{equation}
|\psi (t)\rangle =\sum_{n=0}^{\infty }\sum_{\mathcal{S}(n)}e^{-it\lambda _{n,%
\mathcal{S}}}A_{n,\mathcal{S}}(t)|\varphi _{n,\mathcal{S}}\rangle ~,
\end{equation}%
where $\lambda _{n,\mathcal{S}}$ and $|\varphi _{n,\mathcal{S}}\rangle $ are
the $n$-excitations eigenvalues and eigenstates (\emph{dressed states}) of the
Hamiltonian $\hat{H}_{0}$ and the index $\mathcal{S}$ labels different states
with a fixed number of excitations $n$, which is the quantum number associated to the operator $\hat{N}=\hat{n}+|\mathbf{1}\rangle\langle\mathbf{1}|+2|\mathbf{2}\rangle\langle\mathbf{2}|$. As shown in Sec.\ref{sec2}, the range of values of $\mathcal{S}$
depends on $n$, and we denote such degeneration with $g(n)$. Moreover, the number of excitations in the subspace coincides with the number of photons of the state having the atom in its ground ($| \mathbf{0},n\rangle$).

Following the approach detailed in \cite{jpa,tom}
we propose a change of variables that maps each group of $g(m)$ variables $A_{m,\mathcal{T}}$ into another set $b_{m,\mathcal{T}}$, so that $A_{m,\mathcal{T}}=\sum_{\mathcal{T}'} \alpha_{\mathcal{T}\mathcal{T}'} b_{m,\mathcal{T}'}$. In particular, we consider the following transformation:
\begin{eqnarray}
A_{m,\mathcal{T}} &=&e^{i\Phi _{m,\mathcal{T}}(t)}\left\{ e^{-it\nu _{m,%
\mathcal{T}}}b_{m,\mathcal{T}}(t)\right. \\
&&-\frac{1}{2i}\sum_{\mathcal{S}(m)\neq \mathcal{T}}e^{-it\nu _{m,\mathcal{S}}}b_{m,\mathcal{S}%
}(t)\nonumber\\
&&\times\sum\nolimits_{j}^{%
\prime }\sum_{k=0}^{2}\sum_{L=E,G}  \notag \Upsilon _{m,\mathcal{T},\mathcal{S}}^{L,k,j} \\
&&\times \left. \sum_{r=\pm
}e^{ri\phi _{L,k}^{(j)}}\frac{e^{it(\lambda _{m,\mathcal{T}}-\lambda _{m,%
\mathcal{S}}+r\eta ^{(j)})}-1}{\lambda _{m,\mathcal{T}}-\lambda _{m,\mathcal{%
S}}+r\eta ^{(j)}}\right\}  \notag
\end{eqnarray}%
\begin{eqnarray}
\Phi _{m,\mathcal{T}}(t) &=&\sum\limits_{j}\sum_{k=0}^{2}\sum_{L=E,G}\frac{%
\Upsilon_{m,\mathcal{T},\mathcal{T}}^{L,k,j}}{\eta ^{(j)}} \\
&&\times \left[ \cos (\eta ^{(j)}t+\phi _{L,k}^{(j)})-\cos \phi _{L,k}^{(j)}%
\right] ~,  \notag
\end{eqnarray}%
where we divided the sum in two parts: $\sum\nolimits_{j}^{\prime }$ runs
over \lq fast\rq\ frequencies $\eta ^{(j\prime )}\sim \lambda _{m+2,\mathcal{S}%
}-\lambda _{m,\mathcal{T}}$ and $\sum\nolimits_{j}^{\prime \prime }$ runs
over the \lq slow\rq\ ones $\eta ^{(j\prime \prime )}\sim |\lambda _{m,\mathcal{S}%
}-\lambda _{m,\mathcal{T}}|$. The small frequency shift $\nu _{m,\mathcal{T}%
} $ will be given in Eq. (\ref{nu}) and we introduced constant coefficients ($k=0,1,2$)%
\begin{equation}
\Upsilon_{m,\mathcal{T},\mathcal{S}}^{E,k,j}\equiv \varepsilon
_{E,k}w_{E,k}^{(j)}\langle \varphi _{m,\mathcal{T}}|\hat{\sigma}%
_{k,k}|\varphi _{m,\mathcal{S}}\rangle
\end{equation}%
\begin{equation}
\Upsilon_{m,\mathcal{T},\mathcal{S}}^{G,k,j}\equiv \varepsilon
_{G,k}w_{G,k}^{(j)}\langle \varphi _{m,\mathcal{T}}|(\hat{a}\hat{\sigma}%
_{k+1,k}+\hat{a}^{\dagger }\hat{\sigma}_{k,k+1})|\varphi _{m,\mathcal{S}%
}\rangle ~.
\end{equation}

After substituting $A_{m,\mathcal{T}}$ into the Schr\"{o}dinger equation and systematically
eliminating the rapidly oscillating terms via Rotating Wave Approximation (RWA) \cite{jpa}, to the first order in $%
\varepsilon _{E,k}$ and $\varepsilon _{G,k}$ we obtain the approximate
differential equation for the effective probability amplitude%
\begin{eqnarray}
\dot{b}_{m,\mathcal{T}} &=&\sum_{\mathcal{S}(m)\neq \mathcal{T}}\varsigma
_{m,\mathcal{T},\mathcal{S}}e^{it(\tilde{\lambda}_{m,\mathcal{T}}-\tilde{%
\lambda}_{m,\mathcal{S}})}b_{m,\mathcal{S}}  \label{bmt} \\
&+&\sum\nolimits_{j}^{\prime \prime }\sum_{\mathcal{S}(m)\neq \mathcal{T}%
}\Xi _{m,\mathcal{T},\mathcal{S}}^{(j)}e^{it\varpi _{m,\mathcal{T},\mathcal{S%
}}(|\tilde{\lambda}_{m,\mathcal{T}}-\tilde{\lambda}_{m,\mathcal{S}}|-\eta
^{(j)})}b_{m,\mathcal{S}}  \notag \\
&+&\sum\nolimits_{j}^{\prime }\left[ \sum_{\mathcal{S}(m+2)}\Theta _{m+2,%
\mathcal{T},\mathcal{S}}^{(j)}e^{-it(\tilde{\lambda}_{m+2,\mathcal{S}}-%
\tilde{\lambda}_{m,\mathcal{T}}-\eta ^{(j)})}b_{m+2,\mathcal{S}}\right.
\notag \\
&&\left. -\sum_{\mathcal{S}(m-2)}\Theta _{m,\mathcal{S},\mathcal{T}%
}^{(j)\ast }e^{it(\tilde{\lambda}_{m,\mathcal{T}}-\tilde{\lambda}_{m-2,%
\mathcal{S}}-\eta ^{(j)})}b_{m-2,\mathcal{S}}\right] .  \notag
\end{eqnarray}%
The time-independent transition rates between the dressed states are%
\begin{eqnarray*}
\varsigma _{m,\mathcal{T},\mathcal{S}}
&=&i\sum_{k,l=0}^{1}G_{0,k}G_{0,l}\left\{ \sum_{\mathcal{R}(m+2)}\frac{%
\Lambda _{k,m+2,\mathcal{T},\mathcal{R}}\Lambda _{l,m+2,\mathcal{S},\mathcal{%
R}}}{\lambda _{m+2,\mathcal{R}}-\lambda _{m,\mathcal{S}}}\right. \\
&&\left. -\sum_{\mathcal{R}(m-2)}\frac{\Lambda _{k,m,\mathcal{R},\mathcal{T}%
}\Lambda _{l,m,\mathcal{R},\mathcal{S}}}{\lambda _{m,\mathcal{S}}-\lambda
_{m-2,\mathcal{R}}}\right\}
\end{eqnarray*}%
\begin{equation*}
\Xi _{m,\mathcal{T},\mathcal{S}}^{(j)}=\frac{\varpi _{m,\mathcal{T},\mathcal{%
S}}}{2}\sum_{k=0}^{2}\sum_{L=E,G}\Upsilon _{m,\mathcal{T},\mathcal{S}%
}^{L,k,j}e^{-i\varpi _{m,\mathcal{T},\mathcal{S}}\phi _{L,k}^{(j)}}
\end{equation*}%
\begin{eqnarray}
\Theta _{m+2,\mathcal{T},\mathcal{S}}^{(j)} &=&\sum_{k=0}^{1}\frac{G_{0,k}}{2%
}\left\{ -\frac{\varepsilon _{G,k}^{(j)}\Lambda _{k,m+2,\mathcal{T},\mathcal{%
S}}}{G_{0,k}}\right.  \label{tes} \\
&+&\sum_{l=0}^{2}\sum_{L=E,G}\left[ \sum_{\mathcal{R}(m+2)}\frac{\Lambda
_{k,m+2,\mathcal{T},\mathcal{R}}\Upsilon _{m+2,\mathcal{R},\mathcal{S}%
}^{L,l,j}e^{i\phi _{L,l}^{(j)}}}{\lambda _{m+2,\mathcal{R}}-\lambda _{m+2,%
\mathcal{S}}+\eta ^{(j)}}\right]  \notag \\
&-&\left. \left. \sum_{\mathcal{R}(m)}\frac{\Lambda _{k,m+2,\mathcal{R},%
\mathcal{S}}\Upsilon _{m,\mathcal{T},\mathcal{R}}^{L,l,j}e^{i\phi _{L,l}^{(j)}}}{%
\lambda _{m,\mathcal{T}}-\lambda _{m,\mathcal{R}}+\eta ^{(j)}}\right]
\right\}  \notag
\end{eqnarray}%
\begin{equation}
\Lambda _{k,m+2,\mathcal{T},\mathcal{S}}=\langle \varphi _{m,\mathcal{T}}|%
\hat{a}\hat{\sigma}_{k,k+1}|\varphi _{m+2,\mathcal{S}}\rangle ~.
\end{equation}%
Here $\varpi _{m,\mathcal{T},\mathcal{S}}\equiv \mathrm{sign}(\tilde{\lambda%
}_{m,\mathcal{T}}-\tilde{\lambda}_{m,\mathcal{S}})$ and we introduced the
complex modulation depth $\varepsilon _{l}^{(j)}\equiv \varepsilon
_{l}w_{l}^{(j)}\exp (i\phi _{l}^{(j)})$. Moreover, we defined the corrected
eigenfrequencies
\begin{equation}
\tilde{\lambda}_{m,\mathcal{T}}\equiv \lambda _{m,\mathcal{T}}+\nu _{m,%
\mathcal{T}}+\Delta \nu ,
\end{equation}%
where the correction due to counter-rotating terms reads%
\begin{eqnarray}
\nu _{m,\mathcal{T}} &=&\left[ \sum_{\mathcal{S}(m-2)}\frac{\left(
\sum_{k=0}^{1}G_{0,k}\Lambda _{k,m,\mathcal{S},\mathcal{T}}\right) ^{2}}{%
\lambda _{m,\mathcal{T}}-\lambda _{m-2,\mathcal{S}}}\right.  \label{nu} \\
&&\left. -\sum_{\mathcal{S}(m+2)}\frac{\left( \sum_{k=0}^{1}G_{0,k}\Lambda
_{k,m+2,\mathcal{T},\mathcal{S}}\right) ^{2}}{\lambda _{m+2,\mathcal{S}%
}-\lambda _{m,\mathcal{T}}}\right]  \notag
\end{eqnarray}%
and $\Delta \nu $ denotes the neglected contributions smaller than $\nu _{m,%
\mathcal{T}}$ and the terms of the order $\sim (\Upsilon _{m,\mathcal{T},\mathcal{%
S}}^{L,k,j})^{2}/\omega _{0},(\varepsilon _{G,k}\Lambda _{k,m,\mathcal{S},%
\mathcal{T}})^{2}/\omega _{0}$.

Throughout the derivation of the formula (\ref{bmt}) we have assumed the
constraints%
\begin{equation}
\left\vert \lambda _{m,\mathcal{T}}-\lambda _{m,\mathcal{S}}\right\vert
,|\Upsilon _{m,\mathcal{T},\mathcal{S}}^{L,k,j}|,\left\vert \frac{G_{0,k}\Lambda
_{l,m,\mathcal{S},\mathcal{T}}}{\lambda _{m+2,\mathcal{T}}-\lambda _{m,%
\mathcal{S}}}\right\vert G_{0,l}\ll \omega _{0}  \label{vi}
\end{equation}%
\begin{equation*}
G_{0,k}|\Lambda _{k,m+2,\mathcal{S},\mathcal{T}}|\lesssim \omega _{0}~.
\end{equation*}%
Under these approximations we have $|A_{m,\mathcal{T}}|\approx |b_{m,%
\mathcal{T}}|$, so from Eq. (\ref{bmt}) one can easily infer the evolution
of populations of the dressed states. Besides, the
generalization of our method for $N$-level atoms and second-order effects is straightforward \cite{tom}.

It is worth noting that the occurrence of ADCE is essentially governed by the transition rates $\Theta^{(j)}_{m, {\cal T, S}}$ that couple states belonging to subspaces with different numbers of excitations. Of course the whole dynamics is determined also by the transitions occurring inside each subspace, but the annihilation of (two) excitations is possible only in the presence of non negligible $\Theta$-terms.

\section{Analytical regimes\label{sec2}}

We shall confine ourselves to three different regimes of parameters when the dressed states have simple analytical expressions. With the aid of
these formulas we shall be able to evaluate analytically the coefficients $%
\Theta _{m+2,\mathcal{T},\mathcal{S}}^{(j)}$ in the section \ref{sec3}.

The ground state of $\hat{H}_{0}$ is $|\varphi _{0}\rangle =|\mathbf{0}%
,0\rangle $ and the corresponding eigenenergy is $\lambda _{0}=0$. In this
paper we denote $|\mathbf{k},n\rangle \equiv |\mathbf{k}\rangle
_{atom}\otimes |n\rangle _{field}$, where $\mathbf{k}$ stands for the atomic
level and $n$ stands for the Fock state. Moreover, we define the bare atomic transition
frequencies as%
\begin{eqnarray*}
\Omega _{01} &=&E_{0,1}-E_{0,0}\equiv \omega _{0}-\Delta _{1} \\
\Omega _{12} &=&E_{0,2}-E_{0,1}\equiv \omega _{0}-\Delta _{2}~,
\end{eqnarray*}%
where $\Delta _{1}$ and $\Delta _{2}$ are the bare detunings.

\subsection{Two-level atom (2L)}

We include this case ($G_{0,1}=0$) to compare the advantages and
disadvantages of using qutrits instead of qubits. The exact expressions for $%
m\geq 1$ read

\begin{equation}
\lambda _{m,\pm {D}}=\omega _{0}m-\frac{\Delta _{1}}{2}\pm {D%
}\frac{\beta _{m}}{2}  \label{f4}
\end{equation}%
\begin{equation}
|\varphi _{m,\pm {D}}\rangle =\frac{1}{\sqrt{\beta _{m}}}\left[
\sqrt{\beta _{m,\pm }}|\mathbf{0},m\rangle \pm {D}\sqrt{\beta
_{m,\mp }}|\mathbf{1},m-1\rangle \right] ~,  \label{f5}
\end{equation}%
where $\beta _{m}=\sqrt{\Delta _{1}^{2}+4G_{0,0}^{2}m}$, $\beta _{m,\pm
}=\left( \beta _{m}\pm \left\vert \Delta _{1}\right\vert \right) /2$ and we
introduced the \emph{detuning symbol} ${D}=+1\ $for $\Delta _{1}\geq
0$ and ${D}=-1$ for $\Delta _{1}<0$.

For the qutrits we can use Eqs. (\ref{f4}) -- (\ref{f5}) for the subspace
containing a single excitation, $m=1$; the dressed states with $m\geq 2$
excitations are presented below.

\subsection{Double-resonant regime (RR)}

When both $G_{0,0}$ and $G_{0,1}$ are nonzero, first we consider the special
case when $\Delta _{2}=-\Delta _{1}$, so that we have the double-resonance $%
\Omega _{02}=E_{0,2}-E_{0,0}=2\omega _{0}$. The exact formulas read (for $%
m\geq 2$)%
\begin{equation}
\lambda _{m,0}=m\omega _{0}~,~\lambda _{m,\pm {D}}=m\omega _{0}\pm
{D}\varrho _{m,\mp }  \label{musta}
\end{equation}%
\begin{equation*}
|\varphi _{m,0}\rangle =\mathcal{N}_{m,0}^{-1}\left[ -G_{0,1}\sqrt{m-1}|%
\mathbf{0},m\rangle +\sqrt{m}G_{0,0}|\mathbf{2},m-2\rangle \right]
\end{equation*}%
\begin{eqnarray*}
|\varphi _{m,\pm {D}}\rangle &=&\mathcal{N}_{m,\mp }^{-1}\left[
\sqrt{m}G_{0,0}|\mathbf{0},m\rangle \pm {D}\varrho _{m,\mp }|\mathbf{%
1},m-1\rangle \right. \\
&&\left. +\sqrt{m-1}G_{0,1}|\mathbf{2},m-2\rangle \right] ~,
\end{eqnarray*}%
where we defined%
\begin{equation*}
\varrho _{m}=\sqrt{\Delta _{1}^{2}/4+mG_{0,0}^{2}+\left( m-1\right)
G_{0,1}^{2}}
\end{equation*}%
\begin{equation*}
\varrho _{m,\pm }=\varrho _{m}\pm \left\vert \Delta _{1}\right\vert
/2~,~\varrho _{m,0}=\sqrt{mG_{0,0}^{2}+\left( m-1\right) G_{0,1}^{2}}
\end{equation*}%
\begin{equation*}
\mathcal{N}_{m,0}=\varrho _{m,0}~,~\mathcal{N}_{m,\pm }=\sqrt{2\varrho
_{m}\varrho _{m,\pm }}~.
\end{equation*}%
For example, if $G_{0,1}\sim G_{0,0}$ and $|\Delta _{1}|\gg G_{0,0}\sqrt{n}$
for all relevant values of $n$ we have approximately $|\varphi _{m,-{D}}\rangle \sim |\mathbf{1},m-1\rangle $, $|\varphi _{m,{D}}\rangle
\sim (|\mathbf{0},m\rangle +|\mathbf{2},m-2\rangle )/\sqrt{2}$, while for $%
|\Delta _{1}|\ll G_{0,0},G_{0,1}$ (near the atom--field resonance) we get $|\varphi _{m,\pm {D}%
}\rangle \sim (|\mathbf{0},m\rangle \pm \sqrt{2}|\mathbf{1},m-1\rangle +|%
\mathbf{2},m-2\rangle )/2$.

\subsection{Dispersive regime (DR)\label{prova}}

Now we assume that both the atomic transition frequencies are far-detuned
from the cavity frequency%
\begin{equation}
|\Delta _{1}|,|\Delta _{2}|,|\Delta _{1}+\Delta _{2}|\gg G_{0,0}\sqrt{m}%
,G_{0,1}\sqrt{m-1}.  \label{cri1}
\end{equation}%
\qquad\ From the perturbation theory we obtain to the 4th order in 
$G_{0,0}/\Delta _{1}$ and $G_{0,1}/\Delta _{2}$
\begin{equation*}
\lambda _{m,0}=m\omega _{0}+\delta _{1}m\left[ 1+\frac{G_{0,1}^{2}(m-1)}{%
\Delta _{1}(\Delta _{1}+\Delta _{2})}-\frac{G_{0,0}^{2}m}{\Delta _{1}^{2}}%
\right]
\end{equation*}%
\begin{eqnarray*}
|\varphi _{m,0}\rangle &=&\mathcal{N}_{m,0}^{-1}\left[ |\mathbf{0},m\rangle +%
\frac{\rho _{m,0}G_{0,0}\sqrt{m}}{\Delta _{1}}|\mathbf{1},m-1\rangle \right.
\\
&&\left. +\frac{r_{m,0}G_{0,0}G_{0,1}\sqrt{m(m-1)}}{\Delta _{1}(\Delta
_{1}+\Delta _{2})}|\mathbf{2},m-2\rangle \right]
\end{eqnarray*}%
\begin{eqnarray*}
\lambda _{m,1} &=&m\omega _{0}-\Delta _{1}-\left[ \delta _{1}m-\delta
_{2}(m-1)\right] \\
&&\times \left[ 1-\frac{G_{0,0}^{2}m}{\Delta _{1}^{2}}-\frac{G_{0,1}^{2}(m-1)%
}{\Delta _{2}^{2}}\right]
\end{eqnarray*}%
\begin{eqnarray*}
|\varphi _{m,1}\rangle &=&\mathcal{N}_{m,1}^{-1}\left[ |\mathbf{1}%
,m-1\rangle -\frac{\rho _{m,1}G_{0,0}\sqrt{m}}{\Delta _{1}}|\mathbf{0}%
,n\rangle \right. \\
&&\left. +\frac{r_{m,1}G_{0,1}\sqrt{m-1}}{\Delta _{2}}|\mathbf{2},m-2\rangle %
\right]
\end{eqnarray*}%
\begin{eqnarray*}
\lambda _{m,2} &=&m\omega _{0}-\Delta _{1}-\Delta _{2}-\delta _{2}(m-1) \\
&&\times \left[ 1+\frac{G_{0,0}^{2}m}{\Delta _{2}(\Delta _{1}+\Delta _{2})}-%
\frac{G_{0,1}^{2}(m-1)}{\Delta _{2}^{2}}\right]
\end{eqnarray*}%
\begin{eqnarray*}
|\varphi _{m,2}\rangle &=&\mathcal{N}_{m,2}^{-1}\left[ |\mathbf{2}%
,m-2\rangle -\frac{\rho _{m,2}G_{0,1}\sqrt{m-1}}{\Delta _{2}}|\mathbf{1}%
,m-1\rangle \right. \\
&&\left. +\frac{r_{m,2}G_{0,0}G_{0,1}\sqrt{m(m-1)}}{\Delta _{2}(\Delta
_{1}+\Delta _{2})}|\mathbf{0},m\rangle \right] ,
\end{eqnarray*}%
where we defined the dispersive shifts $\delta _{1}\equiv G_{0,0}^{2}/\Delta
_{1}$ and $\delta _{2}\equiv G_{0,1}^{2}/\Delta _{2}$. We adopted
an intuitive notation in which the second index in $|\varphi_{m,\mathcal{S}}\rangle$ represents the most probable atomic state in a
given dressed state (for example, in the expansion of $|\varphi_{m,0}\rangle$ the bare state $|\mathbf{0},m\rangle$ appears with the highest weight). The parameters $\rho _{m,\mathcal{S}}$, $r_{m,\mathcal{S%
}}$ and $\mathcal{N}_{m,\mathcal{S}}$ are equal to 1 to the first order in $G_{0,0}/\Delta _{1}$, $G_{0,1}/\Delta _{2}$ and are summarized in \cite{SM}.

\subsection{Mixed regime (MR)}

In the mixed regime we assume $\Delta _{2}=0$ and
\begin{equation}
|\Delta _{1}|\gg G_{0,0}\sqrt{n},G_{0,1}\sqrt{n-1},  \label{cri2}
\end{equation}%
i. e., the atomic transition $|\mathbf{1}\rangle \rightarrow |\mathbf{2}%
\rangle $ is resonant with the cavity mode, while the transition $|\mathbf{0}%
\rangle \rightarrow |\mathbf{1}\rangle $ is far-detuned. To the second
order in $G_{0,0}/\Delta _{1}$ we obtain%
\begin{equation*}
\lambda _{m,0}=m\omega _{0}+\frac{\Delta _{1}G_{0,0}^{2}m}{\Delta
_{1}^{2}-G_{0,1}^{2}(m-1)}
\end{equation*}%
\begin{eqnarray*}
|\varphi _{m,0}\rangle &=&\mathcal{N}_{m,0}^{-1}\left\{ G_{0,1}\sqrt{m-1}%
\rho _{m,0}|\mathbf{2},m-2\rangle \right. \\
&&\left. +\rho _{m,0}\Delta _{1}|\mathbf{1},m-1\rangle +|\mathbf{0},m\rangle
\right\}
\end{eqnarray*}%
\begin{eqnarray*}
\lambda _{m,\pm {D}}&=&m\omega _{0}-{D}\left( \left\vert
\Delta _{1}\right\vert \mp G_{0,1}\sqrt{m-1}\right. \\
&&\left.+\frac{1}{2}\frac{G_{0,0}^{2}m}{%
\left\vert \Delta _{1}\right\vert \mp G_{0,1}\sqrt{m-1}}\right)
\end{eqnarray*}%
\begin{eqnarray*}
|\varphi _{m,\pm {D}}\rangle &=&\mathcal{N}_{m,\pm }^{-1}\left\{
(1-r_{m,\pm })|\mathbf{2},m-2\rangle \right. \\
&&\left. \pm {D}(1+r_{m,\pm })|\mathbf{1},m-1\rangle +\rho _{m,\pm }|%
\mathbf{0},m\rangle \right\} ,
\end{eqnarray*}%
where we defined%
\begin{equation*}
\rho _{m,\pm }=\frac{G_{0,0}\sqrt{m}}{G_{0,1}\sqrt{m-1}\mp \left\vert \Delta
_{1}\right\vert }~,~\rho _{m,0}=\frac{G_{0,0}\sqrt{m}}{\Delta
_{1}^{2}-G_{0,1}^{2}(m-1)}
\end{equation*}%
\begin{equation*}
r_{m,\pm }=\frac{1}{4}\frac{G_{0,0}^{2}m}{G_{0,1}\sqrt{m-1}(G_{0,1}\sqrt{m-1}%
\mp \left\vert \Delta _{1}\right\vert )}
\end{equation*}%
\begin{equation*}
\mathcal{N}_{m,0}=\sqrt{1+\rho _{n,0}^{2}\left[ \Delta
_{1}^{2}+(m-1)G_{0,1}^{2}\right] }~
\end{equation*}%
\begin{equation*}
\mathcal{N}_{m,\pm}=\sqrt{2+2r_{m,\pm }^{2}+\rho _{m,\pm }^{2}}\, .
\end{equation*}

\section{ADCE}

\label{sec3}

Our goal is to study the coherent annihilation of system excitations from
the initial separable state $\hat{\rho}_{0}=|\mathbf{0}\rangle \langle
\mathbf{0}|\otimes \hat{\rho}_{th}$, where $\hat{\rho}_{th}=\sum_{m=0}^{%
\infty }\rho _{m}|m\rangle \langle m|$ is the cavity \emph{thermal state}
with $\rho _{m}=\bar{n}^{m}/\left( \bar{n}+1\right) ^{m+1}$. Here $\bar{n}%
=\left( e^{\omega \beta }-1\right) ^{-1}$ is the average initial photon
number, $\beta ^{-1}=k_{B}T$, $T$ is the absolute temperature and $k_{B}$ is
the Boltzmann's constant. From Eq. (\ref{tes}) it is clear that such process
can be implemented via transition of the form $|\varphi _{m,\mathcal{T}%
}\rangle \rightarrow |\varphi _{m-2,\mathcal{S}}\rangle $ when the
modulation frequency is $\eta ^{(res)}= \tilde{\lambda}_{m,\mathcal{T}}-%
\tilde{\lambda}_{m-2,\mathcal{S}}$. So first we must determine the dressed
states for which the initial population of the state $|\varphi _{m,\mathcal{T%
}}\rangle $, denoted as $P_{m,\mathcal{T}}$, is larger than $P_{m-2,\mathcal{%
S}}$. We assume a small integer $m$ (for the sake of illustration we choose $%
m=4$, although the overall behavior is similar for other values of $m$) and
set the realistic parameters $G_{0,0}=6\times 10^{-2}\omega _{0}$ and $\bar{n%
}=1.5$. We verified numerically that when $G_{0,1}$ is of the same order of $%
G_{0,0}$ the exact value of $G_{0,1}$ does not affect qualitatively the
results, so in this paper we set $G_{0,1}=1.2G_{0,0}$. See \cite{SM} for an illustration of the
quantitative differences in the results when $G_{0,1}=G_{0,0}$ or $%
G_{0,1}=0.8G_{0,0}$.

\begin{figure}[tbh]
\centering\includegraphics[width=1.\linewidth]{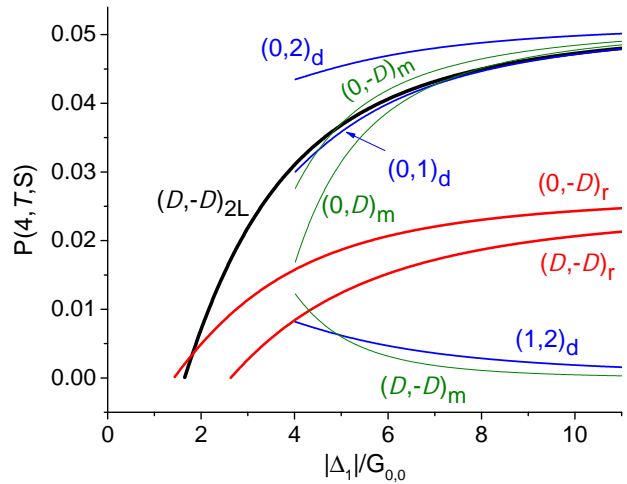}
\caption{(color online) Difference of initial populations $P\left( m,\mathcal{T},%
\mathcal{S}\right) \equiv P_{m,\mathcal{T}}-P_{m-2,\mathcal{S}}$ for $m=4$ and
different regimes as function of the absolute value of the detuning $\Delta
_{1}$. Regimes: 2-level atom (2L), double-resonant regime (r), dispersive
regime\ (d) and mixed regime (m). Only the states for which $P\left( m,%
\mathcal{T},\mathcal{S}\right) >0$ are plotted and the values $(\mathcal{T},%
\mathcal{S})$ are indicated alongside the curves. (Here $G_{0,1}=1.2 G_{0,0}$.)}
\label{figZ1}
\end{figure}

In Fig. \ref{figZ1} we plot the initial population difference $P\left( m,%
\mathcal{T},\mathcal{S}\right) \equiv P_{m,\mathcal{T}}-P_{m-2,\mathcal{S}}$
as function of $|\Delta _{1}|$ for $m=4$. Only positive values of $P\left( 4,%
\mathcal{T},\mathcal{S}\right) $ are plotted and the values $(\mathcal{T},%
\mathcal{S})$ are indicated next to the curves, where the index stands for
2-level (2L), double-resonant (r), dispersive (d) and mixed (m) regimes. In
the dispersive and mixed regimes we assume $|\Delta _{1}|/G_{0,0}\geq 4$ in
order to satisfy the approximations (\ref{cri1}) and (\ref{cri2}). Besides,
throughout this paper we set $\Delta _{2}=6G_{0,0}\mathrm{sign}\left( \Delta
_{1}\right) $ in the dispersive regime so that $|\Delta _{1}+\Delta _{2}|$
never approaches zero, as required by the inequality (\ref{cri1}). One can
see that large detuning $|\Delta _{1}|$ favors the implementation of ADCE;
the transitions $(1,2)_{d}$ and $({D},-{D})_{m}$ are not
particularly useful since the population differences are always small and
are inversely proportional to the detuning. As already known, for a qubit
the ADCE relies on the transition $\left( {D},-{D}\right)
_{2L}$. From Fig. \ref{figZ1} we discover that for a qutrit we have the
following candidates for the realization of ADCE; $\left( {D},-%
{D}\right) _{r}$ and $\left( 0,-{D}\right) _{r}$ in the
double-resonant regime; $\left( 0,1\right) _{d}$ and $\left( 0,2\right) _{d}$
in the dispersive regime; $\left( 0,{D}\right) _{m}$ and $\left( 0,-%
{D}\right) _{m}$ in the mixed regime.

\begin{figure}[tbh]
\centering\includegraphics[width=1.\linewidth]{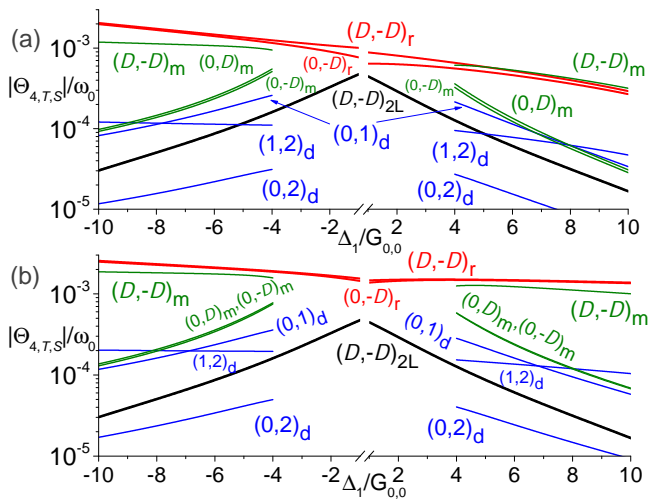}
\caption{(color online) a) Transition rate for ADCE in different regimes as
function of $\Delta _{1}/G_{0,0}$ for $m=4$ and modulation of $E_{1}$. b)
Same as (a) but for the simultaneous modulation of $E_{1}$ and $E_{2}$ with
the same frequency. In the dispersive regime (d) we set $\Delta _{2}=6G_{0,0}%
\mathrm{sign}(\Delta _{1})$. In the mixed regime (m) the lines $(0,{D
})$ and $(0,-{D})$ are very close, so for the sake of compactness
they are not discerned separately. We do not show the transition rate near $%
|\Delta _{1}|=0$, since all the population differences $P\left( m,\mathcal{T}%
,\mathcal{S}\right) $ are negative in this case. Notice the increment by at
least one order of magnitude of the transition rate in the double-resonant
regime (r) [compared to the qubit's case (2L)] for $|\Delta _{1}|\gg
G_{0,0}$.}
\label{figZ2}
\end{figure}

Now we are in position to evaluate the ADCE rate in different regimes
according to Eq. (\ref{tes}). For the transition $|\varphi _{m,\mathcal{T}%
}\rangle \rightarrow |\varphi _{m-2,\mathcal{S}}\rangle $ [denoted as $(%
\emph{T},\emph{S})$] we evaluate analytically $\Theta _{m,\mathcal{T},%
\mathcal{S}}$ under the resonant modulation frequency $\eta ^{(\mathrm{res}%
)}=\tilde{\lambda}_{m,\mathcal{T}}-\tilde{\lambda}_{m-2,\mathcal{S}}$. In
Fig. \ref{figZ2}a we plot the dimensionless transition rate $\left\vert \Theta _{m,%
\mathcal{T},\mathcal{S}}\right\vert /\omega _{0}$ for $m=4$ assuming the
harmonic modulation of $E_{1}$ with perturbative amplitude $\varepsilon
_{E,1}=5\times 10^{-2}\Omega _{01}$. We disregard the region near $\Delta
_{1}=0$ since $P\left( m,\mathcal{T},\mathcal{S}\right) <0$ in this case, so
ADCE does not occur. We observe that for the qutrit in the dispersive or
mixed regimes the transition rates can be slightly higher than for the
qubit; the rate for the transition $({D},-{D})_{m}$ is
substantially higher than for the qubit, however this transition is not
useful for ADCE due to small population difference $P\left( m,{D},-%
{D}\right) $. We also note that in the dispersive regime one can
induce the transition $|\varphi _{m,0}\rangle \rightarrow |\varphi
_{m-2,2}\rangle $ for modulation frequency $\eta ^{(\mathrm{res})}\approx
4\omega _{0}-\Omega _{02}$, that corresponds approximately to the \emph{%
four-photon transition} $|\mathbf{0},m\rangle \rightarrow |\mathbf{2}%
,m-4\rangle $. However the associated transition rate is even smaller than
the ADCE rate for a qubit, hindering practical applications of such
process.

In the dispersive regime the transition rate and the population difference
for the process $|\varphi _{m,0}\rangle \rightarrow |\varphi _{m-2,1}\rangle
$ [denoted as $(0,1)_{d}$ in the figures] is roughly the same as the process
$|\varphi _{m,{D}}\rangle \rightarrow |\varphi _{m-2,-{D}%
}\rangle $ for a qubit [denoted as $({D},-{D})_{2L}$].
Therefore, the behavior of multi-level atoms with respect to ADCE is similar
to the one for a qubit, provided all the transitions are far detuned from
the cavity frequency. Moreover, for the mixed regime and large detuning $%
|\Delta _{1}|$ the population differences $P\left( m,0,{D}\right) $
and $P\left( m,0,-{D}\right) $ are roughly the same as for the
qubit, while the transition rates are several times larger, so the implementation of ADCE would be facilitated.

The \emph{main finding} of the paper is the observation that in the double-resonant
regime the ADCE rate is at least one order of magnitude larger than for the
qubit, and the difference increases for larger $|\Delta _{1}|$, as can be
seen from Fig. \ref{figZ2}a. Besides, in this regime the
population differences $P\left( m,{D},-{D}\right) $ and $%
P\left( m,0,-{D}\right) $ also increase proportionally to $|\Delta
_{1}|$, achieving sufficiently large values for $|\Delta _{1}|\sim 8G_{0,0}$
(see Fig. \ref{figZ1}). Thus, it seems that one could speed up ADCE by at
least one order of magnitude using three-level atoms in the double-resonant
configuration instead of qubits, provided the detuning $|\Delta _{1}|$ is large
enough.

In real circuit QED setups it might be tricky to modulate only one parameter
at a time, while keeping the other parameters constant. So in figure \ref%
{figZ2}b we consider the simultaneous modulation of $E_{1}$ and $E_{2}$
(with the same modulation frequency $\eta ^{(\mathrm{res}%
)}=\tilde{\lambda}_{m,\mathcal{T}}-\tilde{\lambda}_{m-2,\mathcal{S}}$) assuming parameters $\varepsilon
_{E,1}=5\times 10^{-2}\Omega _{01}$, $\varepsilon _{E,2}=5\times
10^{-2}\Omega _{12}$, $\phi _{E,1}=0$ and $\phi _{E,2}=\pi $. Conveniently the
ADCE transition rates increase even more when compared to an isolated
modulation of either $E_{1}$ or $E_{2}$.

In \cite{SM} we illustrate in details the transition
rates and the population differences for different values of $G_{0,1}$ and
isolated modulations of $E_{2}$, $G_{0}$ and $G_{1}$. It is found that the
modulation of $G_{0}$ does not speed up significantly the transition rate in comparison to a qubit, whereas the modulation of $E_{2}$ or $G_{1}$
does increase the transition rate in the double-resonant regime by at least
one order of magnitude. We also
verified that under the simultaneous modulation of all the parameters ($%
E_{1}$, $E_{2}$, $G_{0}$ and $G_{1}$) the total transition rate is still
substantially higher than for a qubit, provided the phases are
properly adjusted. Hence, the simultaneous modulation of several parameters
is not an issue from the experimental point of view, provided one can
manage the phases $\phi _{l}^{(j)}$ corresponding to different modulation components.

\begin{figure}[tbh]
\centering\includegraphics[width=1.\linewidth]{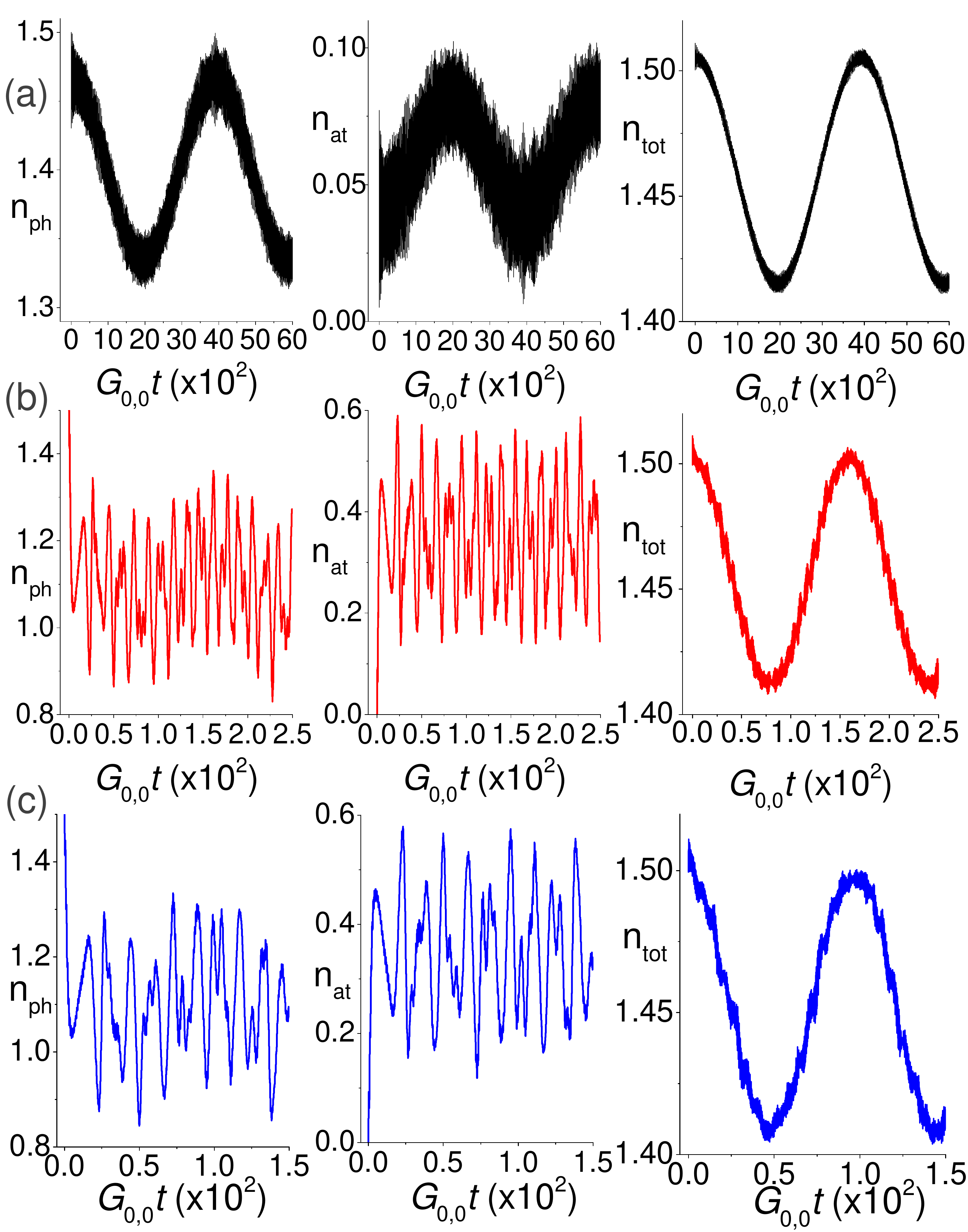}
\caption{(color online) Exact numerical dynamics of ADCE obtained for the
Hamiltonian (\protect\ref{H2}) and the initial local thermal state $\hat{%
\protect\rho}_{0}$ in the double-resonant regime. a) 2-level atom and
harmonic modulation of $E_{1}$. b) 3-level atom and 2-tone modulation of $%
E_{1}$. c) 3-level atom and 2-tone double-modulation of $E_{1}$ and $E_{2}$.
Notice that in all cases the amount of annihilated excitations $n_{tot}$ is
roughly the same, while the duration of the process in (c) is roughly 40
times smaller than in (a).}
\label{figZ3}
\end{figure}
\begin{figure}[tb]
\centering\includegraphics[width=1.\linewidth]{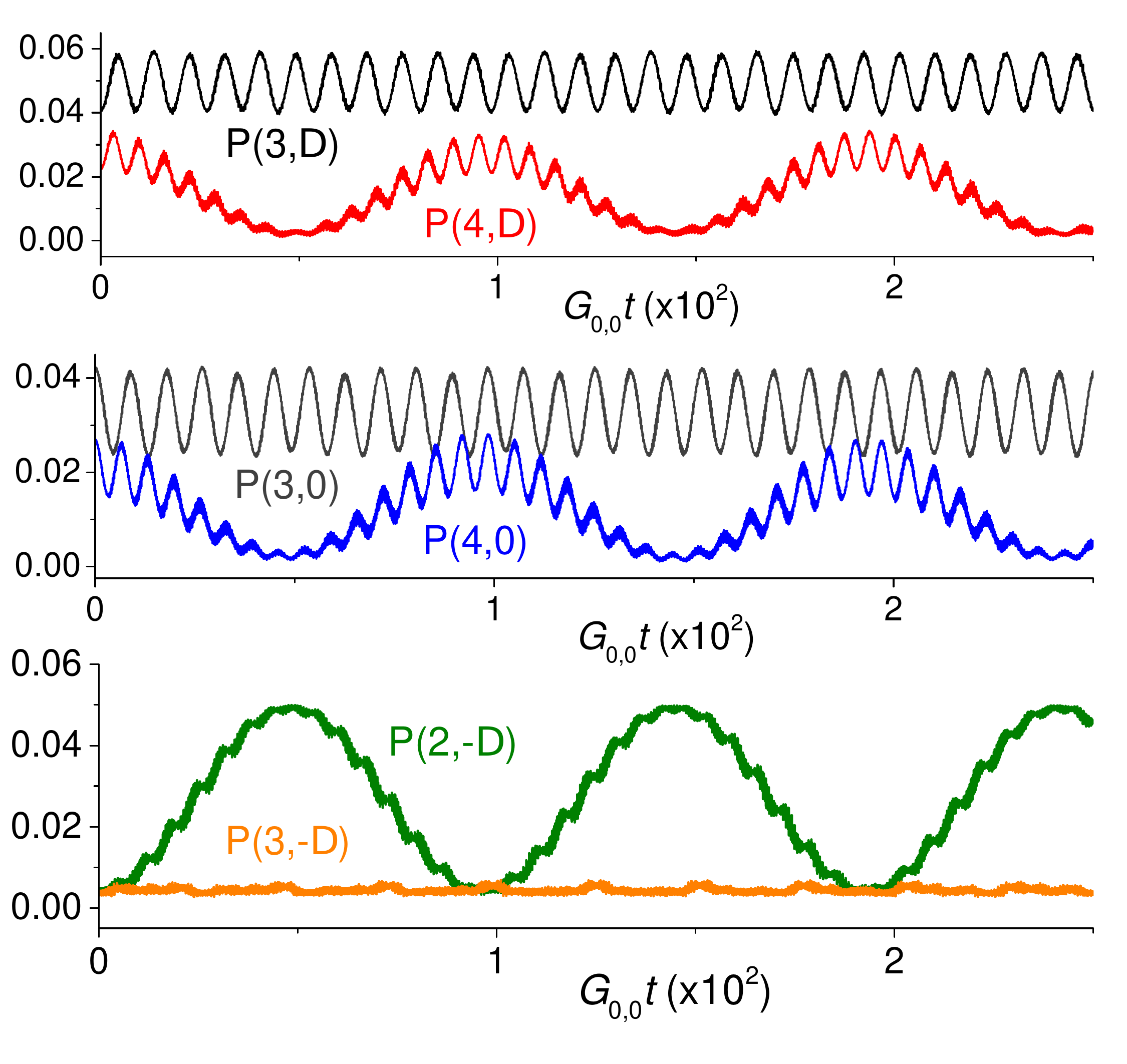}
\caption{(color online) Dynamics of populations of relevant dressed states for
the 2-tone double-modulation of $E_{1}$ and $E_{2}$ analyzed in Fig.
\protect\ref{figZ3}c. There is a coherent transfer of populations from the
states $|\protect\varphi _{4,{D}}\rangle $ and $|\protect\varphi %
_{4,0}\rangle $ to the state $|\protect\varphi _{2,-{D}}\rangle $.
Moreover, one observes periodic oscillations between the dressed states $|%
\protect\varphi _{k,{D}}\rangle \leftrightarrow |\protect\varphi %
_{k,0}\rangle $ due to the counter-rotating terms in the Hamiltonian (%
\protect\ref{H2}).}
\label{figZ4}
\end{figure}

\section{Numerical verification\label{numerical}}

Now we proceed to the numerical verification of the phenomenon predicted in
the previous section, namely, the enhancement of the ADCE rate in the
double-resonant regime. We solved numerically the Schr\"{o}dinger equation
for the Hamiltonian (\ref{H2}) using the initial local thermal state $\hat{%
\rho}_{0}=|\mathbf{0}\rangle \langle \mathbf{0}|\otimes \hat{\rho}_{th}$ and
parameters $m=4$, $G_{0,0}=6\times 10^{-2}\omega _{0}$, $G_{0,1}=1.2G_{0,0}$%
, $\bar{n}=1.5$ and $\Delta _{1}=-\Delta _{2}=-8G_{0,0}$. One downside of
using the double-resonant regime for qutrits is clear from Fig. \ref{figZ1}:
both the populations differences $\left( 0,-{D}\right) _{r}$ and $%
\left( {D},-{D}\right) _{r}$, involved in the ADCE, are
roughly twice smaller than the population difference $\left( {D},-%
{D}\right) _{2L}$ for the qubit. Hence, considering the connection between ADCE and quantum thermodynamic processes recently analyzed in Ref.\cite{tiago}, we can say that the work extraction would be
half smaller if one used qutrits instead of qubits. This nuisance can be readily
surpassed by employing 2-tone modulation with frequencies $\eta ^{(1)}=%
\tilde{\lambda}_{m,0}-\tilde{\lambda}_{m-2,-{D}}$ and $\eta ^{(2)}=%
\tilde{\lambda}_{m,{D}}-\tilde{\lambda}_{m-2,-{D}}$ that
drives simultaneously the transitions $|\varphi _{m,0}\rangle \rightarrow
|\varphi _{m-2,-{D}}\rangle $ and $|\varphi _{m,{D}}\rangle
\rightarrow |\varphi _{m-2,-{D}}\rangle $.

In figure \ref{figZ3}a we illustrate the dynamics of the average photon
number $n_{ph}=\langle \hat{n}\rangle $, the average number of atomic
excitations $n_{at}=\langle \sum_{k=1}^{2}k\hat{\sigma}_{k,k}\rangle $ and
the total average number of excitations $n_{tot}=n_{ph}+n_{at}$ for a
qubit (setting momentarily $G_{1}=0$) with modulation depth $\varepsilon _{E,1}=5\times 10^{-2}\Omega
_{01}$. We observe the sinusoidal oscillation of $n_{ph}$, $n_{at}$ and $%
n_{tot}$ with typical period $\tau \approx 4\times 10^{3}G_{0,0}^{-1}$. The
coherent annihilation of excitations does take place, but since the initial
population of the state $|\varphi _{4,{D}}\rangle $ was $P_{4,%
{D}}\approx 5\times 10^{-2}$, the average number of annihilated
excitations is $\sim 2P_{m,{D}}\approx 0.1$, in agreement with the
numerical data.

In figure \ref{figZ3}b we consider the qutrit under 2-tone modulation
of $E_{1}$ with the previous amplitude $\varepsilon _{E,1}=5\times
10^{-2}\Omega _{01}$, weights $w_{E,1}^{(1)}=10/17,w_{E,1}^{(2)}=7/17$ and
phases $\phi _{E,1}^{(1)}=0$, $\phi _{E,1}^{(2)}=\pi$ (the weights were
adjusted to equalize the two transition rates). We see that the total number
of excitation exhibits the same qualitative behavior as for the qubit, but the transition rate undergoes a 30-fold enhancement. The behavior
of $n_{ph}$ and $n_{at}$ differs drastically from the one observed for the
2-level atom partly due to the oscillations between the bare states $|\mathbf{0}%
,k\rangle \leftrightarrow |\mathbf{2},k-2\rangle $ for $k\geq 2$, and
partly due to the oscillations between the dressed states $|\varphi _{k,%
{D}}\rangle \leftrightarrow |\varphi _{k,0}\rangle $, as will be
discussed shortly. In figure \ref{figZ3}c we consider the simultaneous
two-tone modulation of $E_{1}$ and $E_{2}$ with parameters $\varepsilon
_{E,1}=5\times 10^{-2}\Omega _{01}$, $\varepsilon _{E,2}=9\times
10^{-2}\Omega _{12}$, $w_{E,1}^{(1)}=w_{E,2}^{(1)}=10/17$, $%
w_{E,1}^{(2)}=w_{E,2}^{(2)}=7/17$ and phases $\phi _{E,1}^{(1)}=\phi
_{E,2}^{(2)}=0$, $\phi _{E,1}^{(2)}=\phi _{E,2}^{(1)}=\pi$. We see that
the ADCE rate suffers an additional $50\%$ enhancement compared to
the sole modulation of $E_{1}$, while the average number of total annihilated
excitations is roughly the same as in the previous cases.

Finally, in Fig. \ref{figZ4} we plot the probabilities of finding the system in the dressed states
$P(m,\mathcal{S})={\rm Tr}[\hat{\rho}(t)|\varphi _{m,\mathcal{S}}\rangle \langle
\varphi _{m,\mathcal{S}}|]$ as function of time for the 2-tone
double-modulation discussed in Fig. \ref{figZ3}c. As predicted by Eq. (\ref%
{bmt}) there is a simultaneous periodic transfer of populations from the
states $|\varphi _{4,{D}}\rangle $ and $|\varphi _{4,0}\rangle $ to
the state $|\varphi _{2,-{D}}\rangle $, which correspond to the
coherent annihilation of two system excitations. Other states $|\varphi
_{k\neq 2,-{D}}\rangle $ are not affected by the modulation, as
illustrated for the state $|\varphi _{3,-{D}}\rangle $ which 
undergoes just minor fluctuations due to off-resonant couplings neglected under
RWA. Moreover, one also observes periodic oscillations between the dressed
states $|\varphi _{k,{D}}\rangle \leftrightarrow |\varphi
_{k,0}\rangle $ for $k\geq 2$. This occurs because for large $|\Delta _{1}|$
we have $\tilde{\lambda}_{k,0}\approx \tilde{\lambda}_{k,{D}}$, as
seen from Eq. (\ref{musta}), hence the first term on the RHS of Eq. (\ref%
{bmt}) becomes nearly resonant and couples these states with the strength
$\sim |\varsigma _{k,{D},0}|$ [this behavior is due solely to the counter-rotating terms in Eq. (\ref{H2}) and is independent of modulation].

\section{Conclusions\label{conclusions}}

In conclusion, we showed that the resonant external modulation of
a three-level artificial atom is highly advantageous for the implementation
of the antidynamical Casimir effect (ADCE) in comparison to a two-level atom, since the transition rate can suffer
almost 50-fold increase while the total amount of annihilated excitations is
roughly the same. The strongest enhancement takes place
in the double-resonant regime (when $\Delta _{1}=-\Delta _{2}$, so that $%
\Omega _{02}=2\omega _{0}$) and for large detuning $|\Delta _{1}|$, though weaker
enhancement may occur also in other regimes. Beside speeding up the ADCE,
the use of qutrits also loosens the requirements for accurate tuning of the
modulation frequency, and reproduces the characteristic ADCE behavior of a
qubit when all the atomic transitions are largely detuned from the
cavity field (and $\Omega _{02}\neq 2\omega _{0}$). However, for the optimum annihilation of excitations from a thermal state the usage of qutrits
also brings some inconveniences, such as two-tone
driving and the necessity of controlling the phase difference between different components of the modulation. Nevertheless, our results indicate that the
substantial gain in the transition rate compensates for the additional
complexity in the external control, favoring the experimental implementation
of ADCE.

\begin{acknowledgments}
A. V. D. acknowledges partial support from the Brazilian agency Conselho Nacional de Desenvolvimento Cient\'ifico e Tecnol\'ogico (CNPq).
\end{acknowledgments}

\end{document}